\documentclass[11pt,a4paper,aip,jcp]{revtex4-1}

\usepackage{xcolor}
\usepackage{amsmath}
\usepackage{graphicx}
\usepackage{verbatim}
\usepackage{color}
\usepackage{subfigure}
\usepackage{hyperref}
\begin{document}

\title{Interactions of aqueous amino acids and proteins with the (110) surface of ZnS in molecular dynamics simulations}
\author{Grzegorz Nawrocki and Marek Cieplak}
\affiliation{Institute of Physics, Polish Academy of Sciences, Al. Lotnik\'ow 32/46, 02-668 Warsaw, Poland}

\begin{abstract}
The growing usage of nanoparticles of zinc sulfide as quantum dots
and biosensors calls for a theoretical assessment of interactions of ZnS 
with biomolecules. We employ the molecular-dynamics-based umbrella sampling
method to determine potentials of mean force for 20 single amino acids
near the ZnS (110) surface in aqueous solutions.
We find that five amino acids do not bind at all and the binding energy of the
remaining amino acids does not exceed 4.3 kJ/mol.
Such energies are comparable to those found for ZnO (and to hydrogen
bonds in proteins) but the nature of the specificity is different.
Cysteine can bind with ZnS in a covalent way, $e.g.$ by forming
the disulfide bond with S in the solid. If this effect is included
within a model incorporating the Morse potential, then the potential
well becomes much deeper -- the binding energy is close to 98 kJ/mol.
We then consider tryptophan cage, a protein of 20 residues, and characterize
its events of adsorption to ZnS. We demonstrate the relevance of 
interactions between the amino acids in the selection of optimal
adsorbed conformations and recognize the key role of cysteine in
generation of lasting adsorption. We show that ZnS is more
hydrophobic than ZnO and that the density profile of
water is quite different than that forming near ZnO -- it has only
a minor articulation into layers. Furthermore, the first layer of water
is disordered and mobile.
\end{abstract}

\maketitle

\section{Introduction}

Quantum dots (QD) \cite{ekimov_1981,reed_1988} are semiconducting nanocrystals
in which the exciton-based light emission is tuned by their size \cite{liang_1970}
which allows for simultaneous imaging of many targets \cite{wang_2011}.
They exhibit photonic absorption rates and brightness of the 
emission  that exceed
the corresponding characteristics of organic dyes
by an order of magnitude. These features, together with their
extraordinary photostability, 
suggests using the QDs for biosensing \cite{walling_2009,frasco_2010}. 
However, in order to accomplish sensing of a highly specific nature, the
QDs need to be conjugated with biological molecules such as antibodies, 
which reduces the fluorescence \cite{reiss_2009,wang_2008}
because of a buildup of surface trap states that lead to non-radiative
de-excitation. The novel way out involves 
coating the QDs by a shell of an another semiconductor 
that is characterized by a larger bandgap.
One example of such a core/shell (CS) system
is CdSe/ZnS \cite{dabbousi_1997} in which the core is made
of CdSe and the shell -- of ZnS. ZnS is a semiconductor of a particularly
large gap -- larger than 3.5 eV in the bulk. The precise value 
depend on the crystalline form, temperature and doping.
Coating by ZnS not only improves the fluorescence of the core 
and greatly reduce its toxicity \cite{elbaum_2013}
but it also makes the system stabler in water. Preliminary studies \cite{korczyc_2012}
clearly show solubility of fibers made of ZnO 
and insolubility of the same fibers coated by ZnS shell.
Furthermore, the CS geometry
allows for a wider range of tuning of the emission wavelength \cite{soni_2010,cao_2005}.
Other cores used in the CS systems with ZnS include
CdS, ZnO, and InP \cite{zhu_2006,zhang_2011,brunetti_2013}.
Since ZnS forms the outermost inorganic part of the CS particles 
it is of great interest to understand the nature of interactions
between ZnS and various biomolecules. Here, we present results of all-atom
molecular dynamics (MD) simulations that focus on the strength of interactions
between the solid ZnS and single amino acids (AA) in water solvents.
We then discuss adsorption of small proteins to ZnS.
For simplicity, the ZnS surface is considered to be flat.
This is a necessary first theoretical step to make
before the effects of curvature can be considered.

Our model is described in terms of classical, instead of quantum, 
atoms so that diffusion, conformational transformations and the dynamical effects
of water molecules can be addressed at meaningful time and length scales.
This approach is an accepted standard in simulations of proteins.
Similar models have also been employed
in studies of biomolecules in the vicinity of various solids 
such as Au \cite{hoefling_2010_a,verde_2009}, SiO$_{2}$ \cite{cruz-chu_2006,forte_2007}, 
TiO$_{2}$ \cite{monti_2010}, and ZnO \cite{gao_2009,nawrocki_2013}, but
ZnS appears not yet to be considered. ZnS is, however, different
than ZnO that has been previously studied by us because it may
form covalent bonds with the  sulfur atom of cysteine \cite{sperling_2010,kumar_2013}.
This makes the peptides with multiple cysteines being used as good anchors 
in the process of QDs bioconjugation \cite{bae_1998,pinaud_2004,li_2004}.
One possibility is that the S atoms of ZnS form disulfide bonds with
the S on cysteine. Another is that there is a coordinated binding
by the Zn atoms at the surface imperfections.
Formation and breakage of such covalent bonds is beyond the classical
physics and is not accounted for by just binding the two sulfurs through
the harmonic potential since the resulting forces acts at infinite distances.
Here, we propose a phenomenological way to incorporate the transient
features of the disulfide bond by using the Morse potential.

There are several experimental reports on peptides characterized by a high  
affinity also to other solids such as GaAs, InP and Si
\cite{whaley_2000,bachmann_2010,yokoo_2010,grzyb_2012}. There are also observations
of conformational changes in the peptides induced by adsorption \cite{oren_2005}.
However, the analysis of the interactions with the solid should start
at a more basic level -- that of single amino acids. This step should
help in the interpretation of the adsorption behavior of proteins
and in the design of peptide sequences with a substantial attraction to ZnS.
Here, we follow the approach adopted in the context of ZnO \cite{nawrocki_2013}
and use the umbrella sampling technique \cite{kumar_1992,lemkul_2010} 
to derive the potential of mean force (PMF), $V(z)$, for single AAs in water solutions 
as a function of the distance, $z$, from the surface of ZnS. 
If the resulting $V(z)$ has a minimum then we determine the corresponding binding
energy,  $\epsilon$,  and the bond length, $\sigma$.

We find that five AAs do not bind to ZnS. When they do,
the values of $\epsilon$ are small if one excludes cysteine. Only four
of them exceed 2.1 kJ/mol and are smaller than 4.3 kJ/mol. 
These values are comparable to those obtained for
the four surfaces of ZnO \cite{nawrocki_2013}
and are comparable in strength to the weaker hydrogen bonds in proteins.
Yet, as we shall see here, they are capable of causing brief adsorption
of a small protein without any cysteines. The values of $\epsilon$
obtained for ZnS are comparable but smaller than those calculated 
for  the AA analogues on SiO$_{2}$, 
(up to 12 kJ/mol \cite{wright_2012}) 
and on TiO$_{2}$ (up to 20 kJ/mol \cite{monti_2010}). 
The experimental work of AAs on Si yields values up to
3.64 kJ/mol \cite{basiuk_1996}. Stronger binding
energies, between 17.5 and 44.2 kJ/mol,
have been calculated \cite{hoefling_2010_a} for surfaces of Au.
Unlike the previous examples, all AAs do bind to gold.

Long lasting adsorption of a protein to ZnO can be achieved when
the protein carries a cysteine. We find that the PMF for the
cysteine-ZnS couplings has a well depth of about 100 kJ/mol
which is significantly larger than the values of $\epsilon$
for all other AAs but also substantially smaller than the
dissociation energy of the disulfide bond -- about 264 kJ/mol.
The latter effect reflects the importance of interactions with
the molecules of water.

The PMF derived for AAs through all-atom simulations may be incorporated 
into implicit-solvent C$^{\alpha}$-based coarse grained models of larger
proteins (see, for instance, \cite{pandey_2013a,pandey_2013b}). 
Such simplified models could be used to determine preferred
orientations of adsorption which in turn should enable deduction
of whether the biological functionality of the protein is 
maintained or not. They could also be used to estimate surface
coverage area.

\section{Methods}

Our all-atom MD simulations employ the GROMACS 4.0.7 package \cite{spoel_2005} 
with the AMBER-99 force field \cite{duan_2003}. 
Water molecules are described by the TIP3P model \cite{jorgensen_1983}. 
The interactions between atoms can be subdivided into bonded and non-bonded. 
The latter account for Coulombic interactions between charged particles,
Lennard-Jones-like van der Waals forces and similarly described
repulsion due to overlapping electron orbitals. The bonded interactions
account for stretching of covalent bonds, twisting of
bond angle, and  bending of the dihedral angle. 
We model the dihedral terms by the four-body
Ryckaert-Bellemans potential
\begin{equation}
V_{rb}(\phi_{ijkl})=\sum_{n=0}^{5}C_{n}(cos(\phi_{ijkl}-180^{\circ}))^{n},
\end{equation}
where $\phi_{ijkl}$ is the angle between the $ijk$ and $jkl$ planes.

The covalent bonds are commonly described by the harmonic potential.
So are the bending terms. However, in order 
to render the transient nature of the S-S covalent bond
between the sulfur atom on cysteine and the nearest sulfur atom of ZnS,
we describe such bond by the Morse potential 
\begin{equation}
V_{Morse}(r_{ij})=D_{ij}[1-exp(-\beta_{ij}(r_{ij}-b_{ij}))]^{2} \;\;
\end{equation}
and neglect dissociation of H from the thiol group into the solvent.
Here $D_{ij}$ denotes the depth of the potential well, $\beta_{ij}$ defines
its steepness, and $b_{ij}$ is the location of the minimum.
For small deviations from the minimum, the Morse potential acts like the
harmonic potential  $V_{Morse}(r_{ij})\approx\frac{1}{2}k_{ij}(r_{ij}-b_{ij})^2$.
The corresponding stiffness coefficient, $k_{ij}$ is listed in the
AMBER-99 force field. Parameter $D_{ij}$ is equal to the dissociation energy
\cite{lide_2009} which is added to the list of the force field parameters.
For instance, for the C-N bond $D_{ij}$ is equal to 305 kJ/mol.
Parameter $\beta_{ij}$ is equal to $\sqrt{\frac{k_{ij}}{2D_{ij}}}$.
For reasons of computational convenience, we use the Morse potential
also to describe the remaining covalent bonds -- the peptide bonds, albeit
with a higher $D_{ij}$. 
Furthermore, we use the normalized and sign-inverted
Morse potential as an envelope that multiplies the bond and dihedral potentials 
involving the two atoms linked covalently (such as in the S-S-C$_{\beta}$
triplet in the bond-angle term in the disulfide bond and in the
quadruplet S-S-C$_{\beta}$-C$_{\alpha}$ in the dihedral term).

The stablest crystal form of ZnS -- the zinc blende -- is cubic
and the lattice constant $a$ is equal to 0.541 nm \cite{swanson_1953}. 
The commonest cleavage face is found to be along the (110) plane \cite{tasker_1979}
and this is the face we consider here. 
We model the ZnS surface by a slab of 5 $\times$ 7 $\times$ 2 unit cells.
The solid is considered to be rigid and its geometry is taken to be bulk-like.
The slab is located just under the bottom of the $L_x \times L_y \times L_z$ 
simulation box where water molecules and amino acids (or a protein) are
placed. Here $L_x$ and $L_y$ are about 3.8 nm and are fited to 
the size of the slab. In proper runs, $L_z$ is equal to 4.0 nm were 
the reflecting wall is placed.
Above this wall, there is an empty space extending to $z$ = 3 $\times$ $L_z$.
Another reflecting wall, affecting only the amino acids, is placed at $z$ = 3.5 nm
-- otherwise an AA or a protein might get trapped at the water-vacuum interface.
The purpose of this construction with the vacuum is to allow for the usage of 
the periodic boundary conditions
(in the periodic image, above the empty space there are atoms of the solid)
with the pseudo-2$D$ particle mesh Ewald summation \cite{essman_1995}.

The partial charges on the Zn and S 
atoms in the solid are set to 0.48 $e$ and -0.48 $e$, respectively,
in agreement with the Mulliken analysis \cite{xie_2010}.
Lennard-Jones parameters for both atoms are adopted from AMBER-99 force field. 
The standard procedure of making the simulations feasible in an acceptable time 
is to apply cutoff radii for all relevant interactions. 
We use the cutoff of 1.0 nm combined with the gradual switching off of the 
interactions between 1.0 and 1.2 nm. 

The MD simulations were performed using the leap-frog algorithm 
with a time step of 1 fs. 
Temperature coupling with a Berendsen thermostat at the temperature of
300 K was implemented with the time constant of 0.1 ps.  At the beginning of
each run, the energy of the system was minimized through the steepest descent 
algorithm and the initial velocities were Maxwellian. The first runs pertained
to the situation without any biomolecules -- 
the system was equilibrated for 4 ns with $L_z$ set at 5 nm. 
At that stage $L_z$ was reduced to 4 nm. The reason for the reduction in $L_z$
is the depletion of water at the top due to the attraction by the solid.
Further equilibration continued for 1 ns and only then
the biomolecules and Na$^+$ and Cl$^-$ ions were inserted adiabatically. 
The concentration of the ions corresponded to the physiological 150 mM: 
4 ions of Na$^+$ and 4 ions of Cl$^-$. 
If a biomolecule had a net charge due to the side groups, 
extra ions were inserted to neutralize it. 

The AAs considered in the simulations are endowed with the 
acetyl group at the N-terminus and N-methylamide group at the C-terminus.
These caps eliminate the terminal charges and mimic the presence of a peptide 
chain in which the AA exists in the unionized form 
(see, $e.g.$ \cite{hoefling_2010_a,dragneva_2013}).
Histidine is considered in its three possible protonation states: 
HIE (H on the $\epsilon$ N atom), HID (H on the $\delta$ N atom)
and positively charged HIP (H on both $\epsilon$ and $\delta$ N atoms).
At the assumed value of pH of 7, all three forms are present 
with the same probability.
The VMD software was used for viewing and analyzing the MD results. 

The main objective of the simulations is to determine the PMF for twenty
capped AAs and for the tryptophane cage. The PMF is defined as an
effective potential that corresponds to the average force \cite{kirkwood_1935}
and it is associated with the center of mass (CM) of the molecule
(of the amino acid without the caps).
We determine it by implementing the umbrella sampling method \cite{kumar_1992,lemkul_2010}. 
It involves two stages. In the first stage, one generates 
a set of initial conformations for representative values of $z$ 
by pulling the CM of the object along the $z$-axis -- perpendicular to the
surface of the solid.
Pulling is implemented through a "dummy particle" which moves 
towards the surface with a constant speed of 1 nm/ns from $z$=2 nm to $z$=0 
and drags the CM by the harmonic force  corresponding to the
the spring constant of 5000 kJ/(mol nm$^{2}$). 
The lateral motion is not constrained so the PMF is averaged laterally.
The conformations are scanned every 0.1 ps in order to save them 
with the CM within each of the interval of width 0.05 nm. 
In this way, between 36 and 38 conformations are collected in the case of the AAs
and between 30 and 31 in the case of the protein. 
These conformations are used in the second stage of further runs 
lasting for 40 ns each
(1 ns corresponds to equilibration). In the second-stage runs, 
the $z$-location of the pulling particle is fixed 
and the CM moves within a sampling window of width $\Delta z$. 
The distribution of the resulting vertical locations of CM in the window 
has a maximum where the harmonic pulling force is balancing all forces 
acting on the AA (we exclude the forces acting on the caps) 
or the protein along the $z$ direction. 
This force is averaged over time and locations within each window 
and is then integrated over $z$ to get the PMF. 
The errors of the average forces are determined by a block averaging method \cite{hess_2002} 
and are propagated during integration. 
The standard deviations of average $z$ values within the umbrella sampling
intervals are found to be negligible. 

\section{Results and Discussion}


\subsection{Properties of water near the solid}

\begin{figure}[ht]
\begin{center}
\includegraphics[scale=0.3]{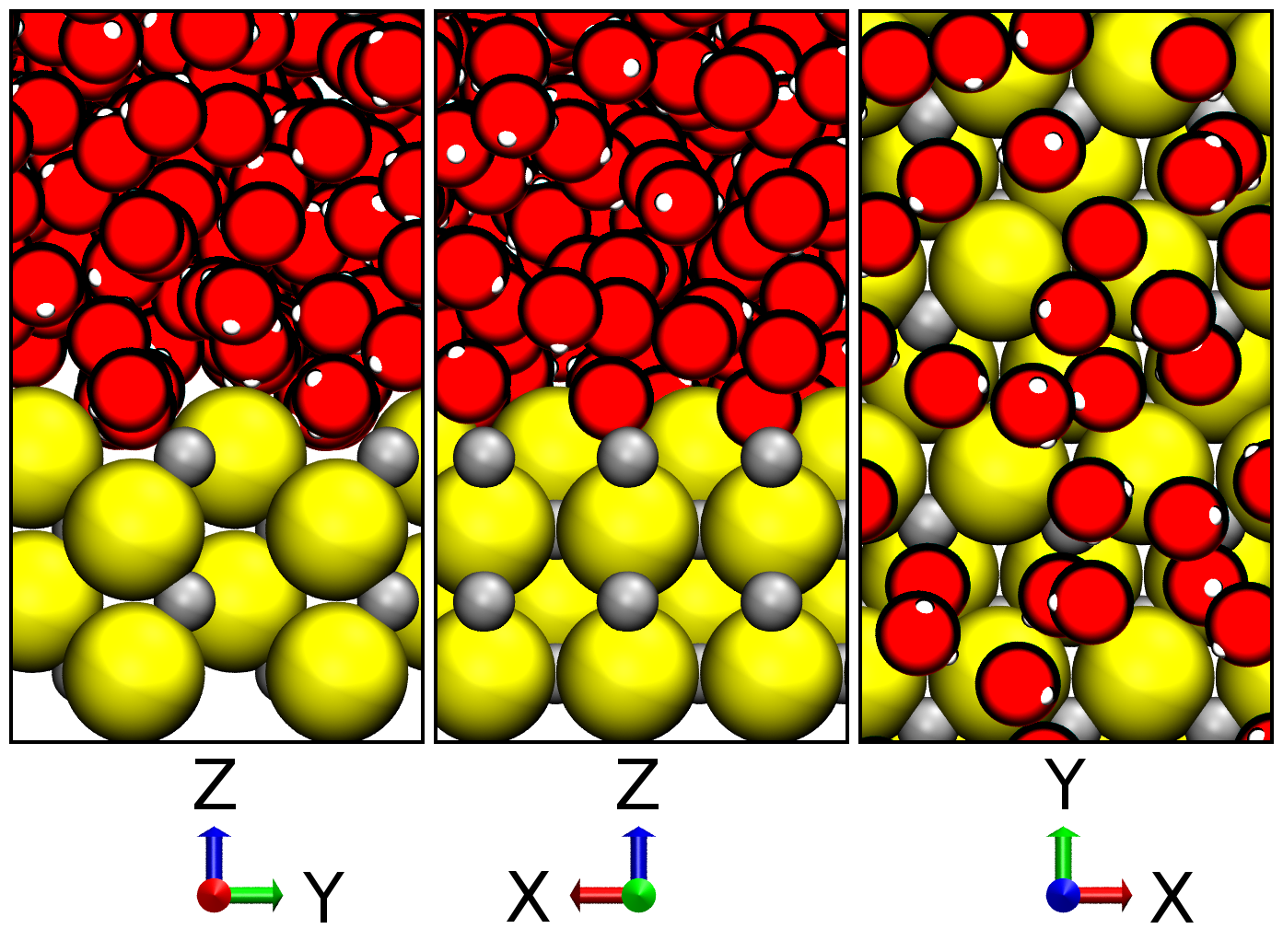}
\end{center}
  \caption{Snapshots of the solid-water interface for the ZnS (110) surface at various projections.
    The $z$ direction is perpendicular to the solid.
    The symbols without the thick black coats show atoms of the solid: 
    the Zn atoms are smaller and are in gray; the S atoms in yellow.
    The symbols with the coat show the atoms of water  (only those
    that are close to the solid are shown in the rightmost panel):
    the O atoms are in red and the H atoms in white. Except for the H atoms,
    the relative sizes of spheres correspond to the distance to
    the minima of the Lennard-Jones potentials.} 
  \label{sur_sol}
\end{figure}

The properties of water, such as the profiles of density and polarization
are derived from a 1 ns simulation performed just after equilibration.
Fig. \ref{sur_sol} shows a snapshot of the solid-water interface. It indicates
that the atomic-level roughness of ZnS allows for a partial penetration by water.
The layer of water mediately next to the solid is disordered which is in contrast
to the nearly frozen character of such a layer near 
ZnO \cite{nawrocki_2013,zhang_2007}. This observation suggests that ZnS should
be more hydrophobic than ZnO and thus also less soluble in water which is
consistent with the experimental findings \cite{elbaum_2013}.

\begin{figure}[ht]
\begin{center}
\includegraphics[scale=0.5]{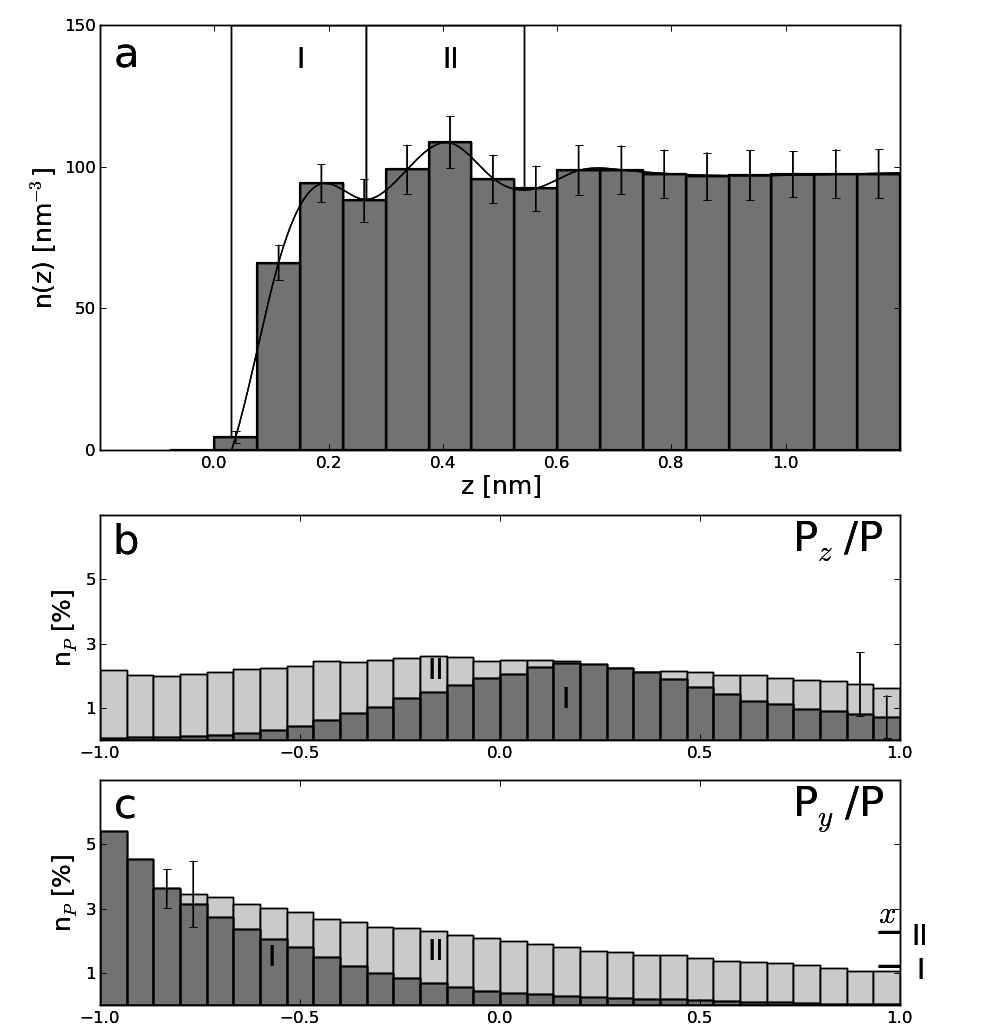}
\caption{a) The number density profile of water molecules above the surface of ZnS. 
         The $z$ coordinate is measured from the center of the topmost atoms of the solid.
         b) The distributions of water polarization for the first two water layers 
           - the darker gray color corresponds to the first layer
           and the lighter to the second. $P$ is the magnitude of the polarization
           vector and $P_{z}$ denotes its $z$-component. 
           $n_P$ is the number of water molecules with a given polarization 
           divided by the number of all water molecules in the two layers
           and expressed as a percentage.
           The contents of the bins in the two layers add up together to 100\%.
           The average errors for both layers are shown in selected bins.
         c) Similar to b) but for the $y$ component of the vector of
         polarization. The $x$-component is approximately uniform and it hovers
         around the levels indicated by two horizontal bars on the right
         for the first and second layers respectively.} 
\label{sol_den_pol}
\end{center}
\end{figure}

The number density profile is shown in panel a of Fig. \ref{sol_den_pol}. It is seen to be 
very poorly articulated which is quite distinct from what has been
derived for other solid surfaces such as ZnO
 \cite{nawrocki_2013}, Au \cite{hoefling_2010_a,verde_2009,chang_2008}, 
SiO$_{2}$ \cite{castrillon_2009} and TiO$_{2}$ \cite{monti_2010}
where patterns of maxima and minima are observed.
The lack of pronounced articulation is a result of rather weak
attraction of water molecules to ZnS. 
Furthermore, there are more molecules in the second layer compared to the
first one which is also different from the density profiling found
in model noble gases near repulsive walls \cite{cieplak_1999,koplik_2000}.
We attribute this difference to the atomic roughness, $i.e.$ to the
size differences between the atoms of Zn, S, and O.
The two bottom panels of Fig. \ref{sol_den_pol} show the profiles
of the normalized $y$ and $z$-components of the water polarization vector.
The distribution for the $x$ component is nearly uniform indicating
lack of polarization in this direction. In both of these panels,
the distribution for the first water layer is less uniform than for the second 
pointing to an ordering influence of the solid in its closest proximity. 
The profile of the  $z$-component shows that maxima for both layers are 
fairly symmetrical with respect to the zero value 
suggesting that the polarization of the first layer 
causes the polarization of the second. 
The profile of the $y$-component shows consistent maxima for both layers 
indicating both of them are polarized by the surface in this direction. 
The polarization in the $z$-direction is more disordered and more uniform
than in the $y$  direction.


\subsection{PMF for amino acids}

\begin{figure}[h!]
\centering
\subfigure[]{
\includegraphics[scale=0.5]{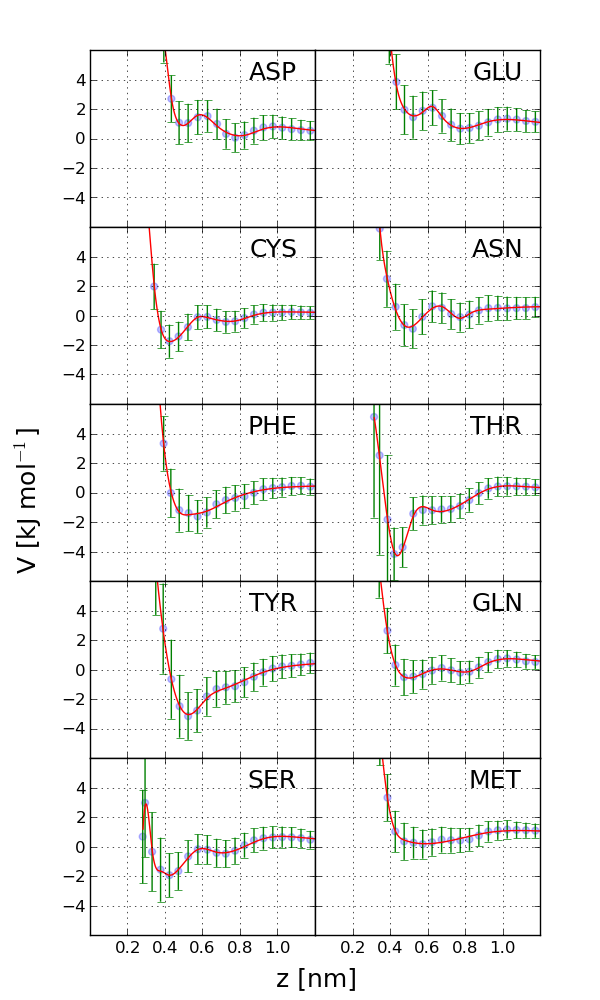}}
\subfigure[]{
\includegraphics[scale=0.5]{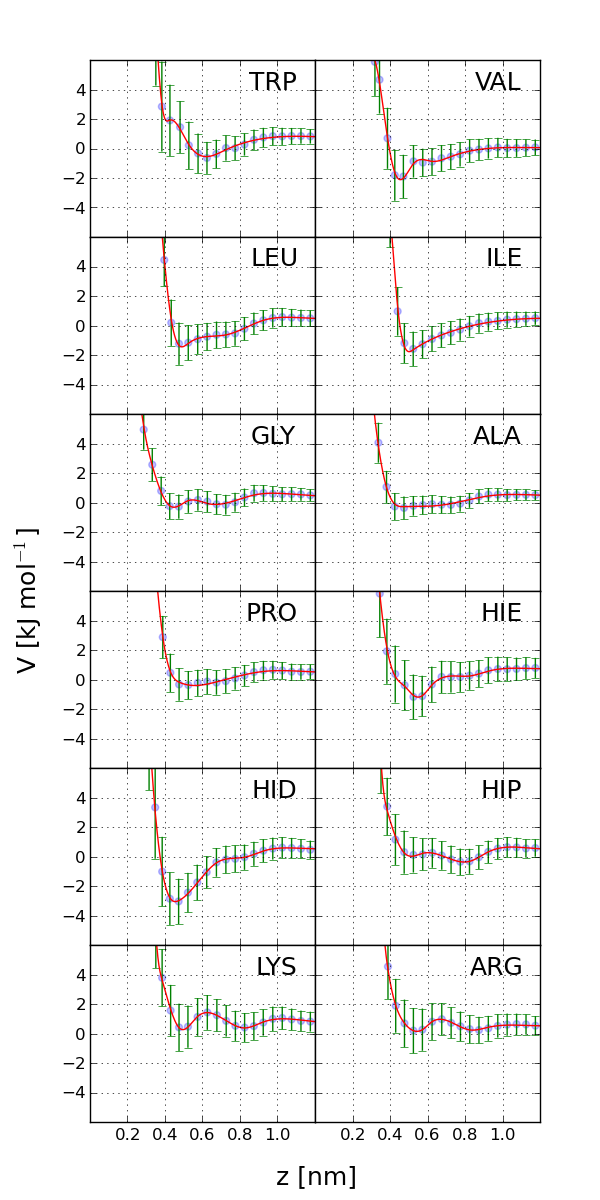}}
\caption{The potentials of the mean force (PMF) for single capped AAs in water solutions 
                 as a function of the distance $z$ above the surface of ZnS.
                 The depth of the lowest negative minimum in $V(z)$ defines the binding energy, $\epsilon$.}
\label{aas_pmf}
\end{figure}

Fig. \ref{aas_pmf} shows the PMFs for the twenty capped AAs.
Histidine is considered in its three states of protonation. The potentials
$V(z)$ are repulsive within the first 0.4 nm away from the surface. 
The potentials for ASP, GLU, MET, LYS, and ARG continue to be
repulsive at larger distances.
The remaining amino acids are weakly attracted within up to 1.2 nm.
Some of the plots of PMFs show undulations. This happens in the case of
the charged amino acids -- ASP, GLU, ARG, and LYS, but not HIS --
and in the case of AAs with a small polar groups -- ASP and THR.
Generally, undulations are due to layering in the density profiles
of water \cite{nawrocki_2013} but the lack of strong features in the profile
makes the undulations weak for ZnS. The values of $\epsilon$ and
$\sigma$ for the binding situations are listed in Table \ref{aas_val}.
Some of these values are small and comparable to the
error bars -- the case of GLY, ALA, PRO, and HIP. These AAs are
essentially neutral to the surface.
The strongest binding is observed for THR, TYR, and HID.
The corresponding values of $\epsilon$ are between 3.06 and 4.28 kJ/mol.

Our results can be compared to those obtained by 
Wright $et$ $al.$ \cite{wright_2012} 
for AA analogues at hydroxylated SiO$_{2}$ surface and 
to those obtained by Monti $et$ $al.$ \cite{monti_2010} for such
analogues at the surface of TiO$_{2}$. 
The binding energies for these two other systems are in the same
range of what we obtain for ZnS and much smaller than
what has been reported for gold (17.5-44.2 kJ/mol as found by 
Hoefling $et$ $al.$ \cite{hoefling_2010_a} and 3-22 kcal/mol as found by 
Feng $et$ $al.$ \cite{feng_2010}).
In the case of SiO$_{2}$,
the behavior is similar because the polar and hydrophobic analogues adsorb 
at the surface whereas positively charged ones do not. It is
also different because the negatively charged analogues bind
to SiO$_{2}$ but not to ZnS.
In the case of TiO$_{2}$, the hydrophobic analogues show poor binding
and all charged moieties bind well.

\begin{figure}[ht]
\begin{center}
\includegraphics[scale=0.4]{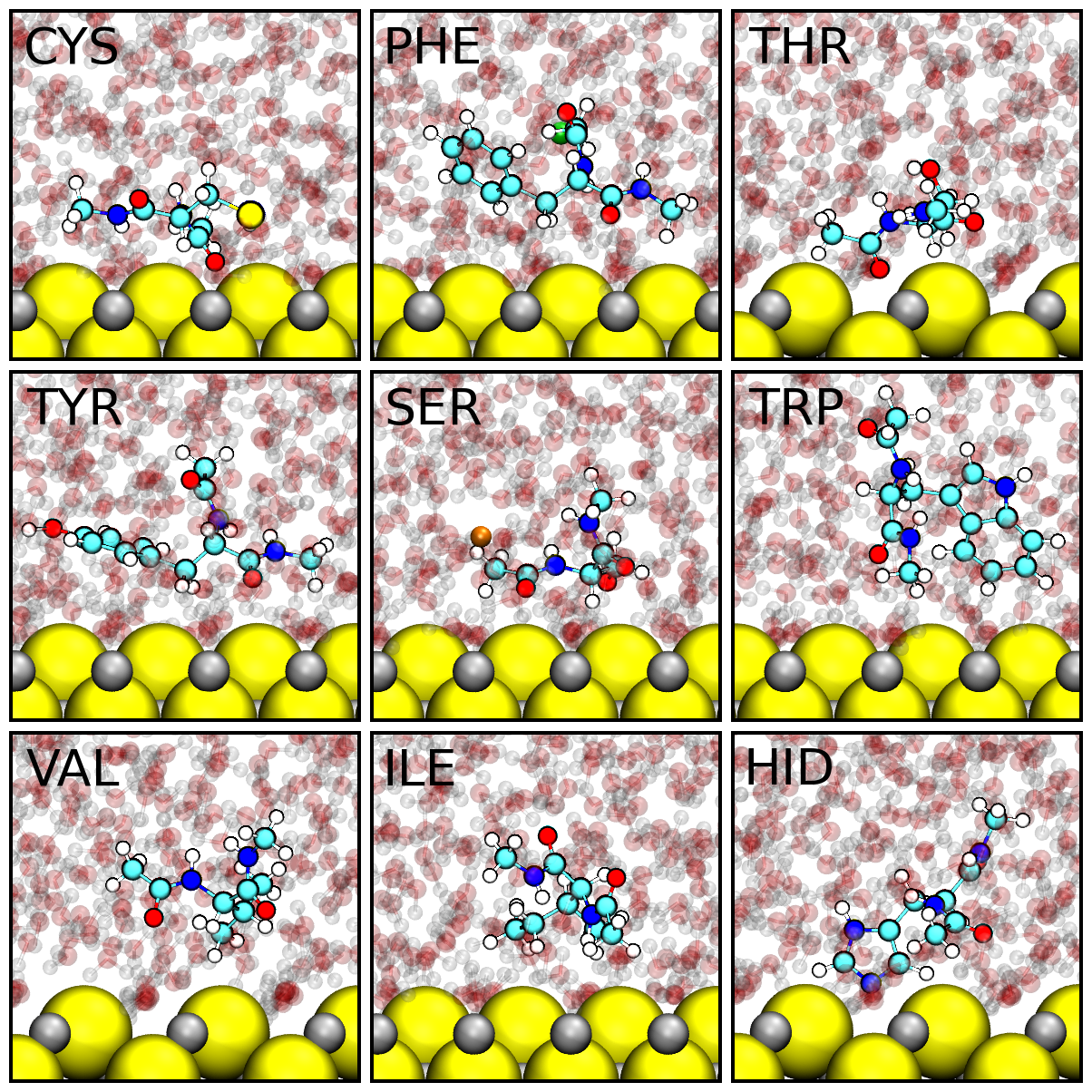}
\end{center}
        \caption{The optimal conformations of the nine AAs 
  corresponding to the highest binding energy to the surface of ZnS.}
\label{aas_con}
\end{figure}

Fig. \ref{aas_con} shows examples of conformations for nine AAs as 
obtained at the minimum of the PMF. 
The AAs are seen to interact
directly with the surface which is distinct from interactions
through the first (and well articulated) layer of water that characterized
interactions with ZnO \cite{nawrocki_2013}. 
The screening by water weakens the binding and yet the absence
of this effect in the case of ZnS yields comparable values of $\epsilon$.
There are two reasons for this behavior. The first is that the partial charges
of atoms in ZnS  are smaller than those in ZnO by about a factor of 2. 
The second is that even though the water molecules do
not form a frozen layer (as the ZnS surface is
less hydrophilic than ZnO), they do compete with most of the AAs
for the access to the surface of ZnS and weaken the attachment.  

The conformations shown in Fig. \ref{aas_con} indicate that the binding
takes place when the O and H atoms either on the side chain or on the
peptide bonds substitute the O and H atoms of the water molecules
in their direct interaction with ZnS. The substitution of O is
seen in the case of CYS and THR, whereas the substitution of H
-- in the case of TRP and ILE. 
The adsorption of HID is caused by the substitution of O by the N in the
$\epsilon$ position of the ring which is not the protonated N
in the $\delta$ position.
The binding energy of HIE is about three times smaller than that of HID.
The reason for the weaker coupling is that in HIE the protonated N is in the 
$\epsilon$ position and the unprotonated N in the $\delta$ position
and the $\delta$ position is less exposed to the surface of ZnS.
We conclude that the nature of conformation at the minimum of PMF
affects the binding energy in a substantial way.

We should point out that the conformations shown in Fig. \ref{aas_con}
do not include the conformations of the hydrophobic methyl groups (CH$_3$) 
on the caps. The methyl groups are known to be vital in adsorption to Au
as they appears to initiate the event \cite{hoefling_2010_a}.
In the case of ZnS, however, these groups appear to play only a minor role.

\subsection{Interactions with cysteine}

\begin{figure}[ht]
\begin{center}
\includegraphics[scale=0.5]{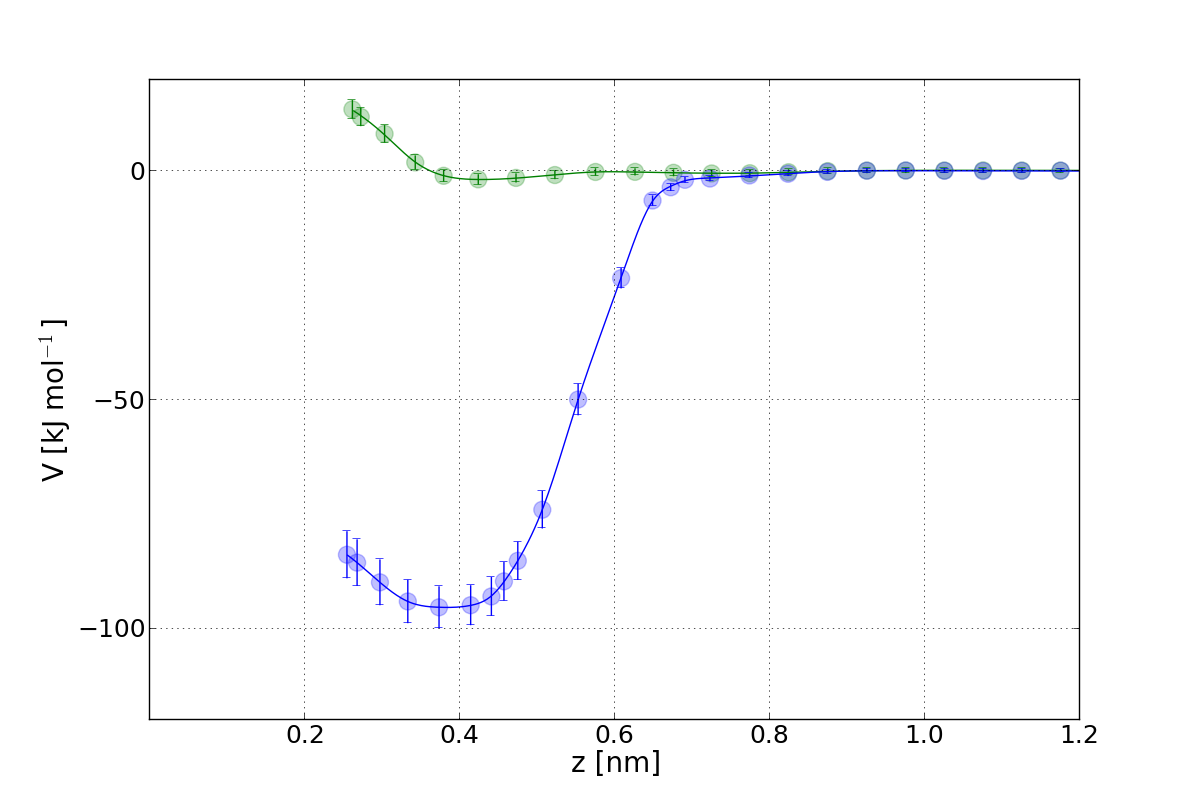}
\caption{The PMF for CYS in water solution 
         as a function of the distance $z$ above the surface of ZnS. 
         The data points at the top (in green) correspond to non-covalent binding
         (a more detailed view is given in Fig. \ref{aas_pmf}).
         The data points at the bottom (in blue) are obtained for mCYS, $i.e.$ when the
         model disulfide coupling is included.}
\label{cys_pmf}
\end{center}
\end{figure}

\begin{figure}[ht]
\begin{center}
\includegraphics[scale=0.3]{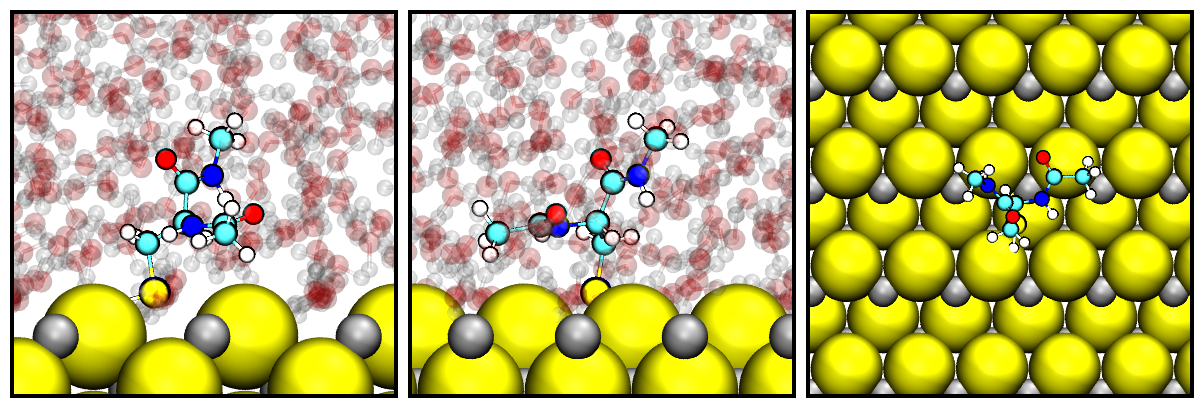}
\end{center}
        \caption{The optimal conformation of mCYS adsorbed at the surface of ZnS.
          The three panels correspond to three projections on the $yz$, $xz$, and $xy$
           planes respectively.}
\label{cyss_con}
\end{figure}

In our analysis so far, CYS has had no capability of forming a covalent bond.
We now consider a situation in which CYS forms both non-covalent
and disulfide bonds. This will be denoted as mCYS -- for modified CYS.
Fig. \ref{cys_pmf} compares $V(z)$ obtained for CYS and mCYS. 
It is seen that the replacement of CYS by mCYS increases $\epsilon$
nearly 50 times while it decreases $\sigma$ only slightly.
The widths of the potential wells stay about the same.
$V(z)$ shows undulations for CYS but no undulations for mCYS
which suggests a smaller sensitivity to the density profiling of water.
Fig. \ref{cyss_con} shows the optimal conformation of mCYS 
bonded covalently with an S atom of ZnS -- it is quite distinct
to the non-covalent binding shown in panel a of Fig. \ref{aas_con}.

The PMF has been obtained within the scheme of the umbrella sampling.
We now consider MD without any steering elements in which either CYS or mCYS
diffuse around. 
Over 13 \% of the entire time of our simulation, is recognized 
as corresponding to adsorption of CYS which means that
at least one of the CYS atoms is within 0.5 nm of the surface. 
Replacing CYS by mCYS increases the fraction of adsorption time to 82 \%.
In practice, once mCYS touches the surface it remains adsorbed for the rest  
of the simulations, however, spontaneous desorption may still occur. 
Thus our modeling of the covalent binding is fairly adequate.

\begin{figure}[ht]
\begin{center}
\includegraphics[scale=0.5]{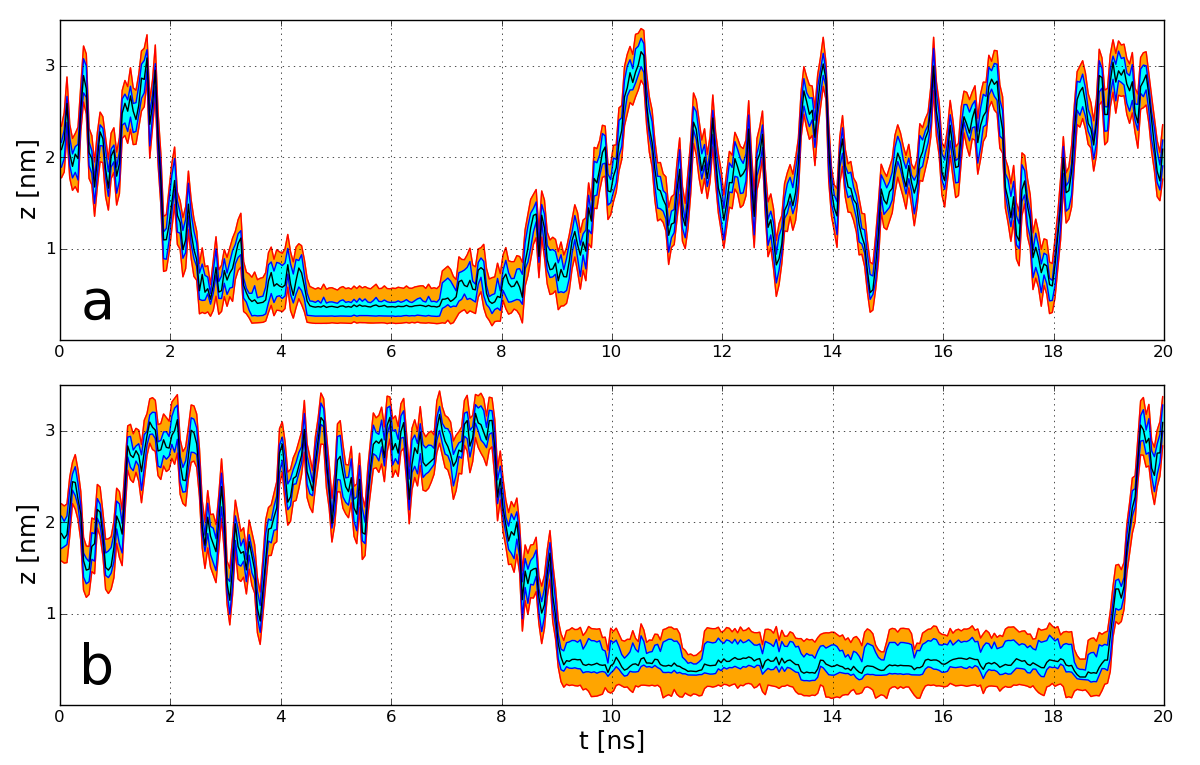}
\end{center}
\caption{The free evolution of cysteine in water near the surface of ZnS. 
The starting location is at 2 nm above the surface.
In panel a) CYS is considered to be
forming solely noncovalent bond with the surface.
Panel b) corresponds to mCYS, $i.e.$
CYS is considered capable of forming both noncovalent and covalent bonds.
The lines show, top to bottom, the instantaneous vertical positions of: the
highest atom, the highest CM of the three groups -- amino acid and two caps, 
the CM of the whole molecule (with the caps), the CM of the lowest group and the lowest atom.}
\label{cys_dyn}
\end{figure}

It is interesting to note that,
in the absence of the disulfide bond, 
the conformation of adsorbed cysteine is stable and its orientation 
with respect to the surface is persistent (see Fig. \ref{cys_dyn}). 
This is because CYS adheres to the surface with many atoms simultaneously, 
as shown in Fig. \ref{aas_con}.
With the covalent bond, however, the distances of the atoms from the
surface are found to fluctuate more vigorously (the bottom panel of
Fig. \ref{cys_dyn}) and the conformations are more dynamic 
in nature. The reason for the different 
behavior is that the formation of the covalent  bond induces a
conformation, shown in Fig. \ref{cyss_con}, in which other 
parts of the molecule are further away from the solid. Thus the
covalently bound molecule is more prone to the influence of water. 
This influence affects the optimal bonding length and angles
and encourages desorption. More stabilization is expected to take 
place when CYS is a part of a protein.

\subsection{Tryptophane cage near ZnS}

\begin{figure}[ht]
\begin{center}
\includegraphics[scale=0.5]{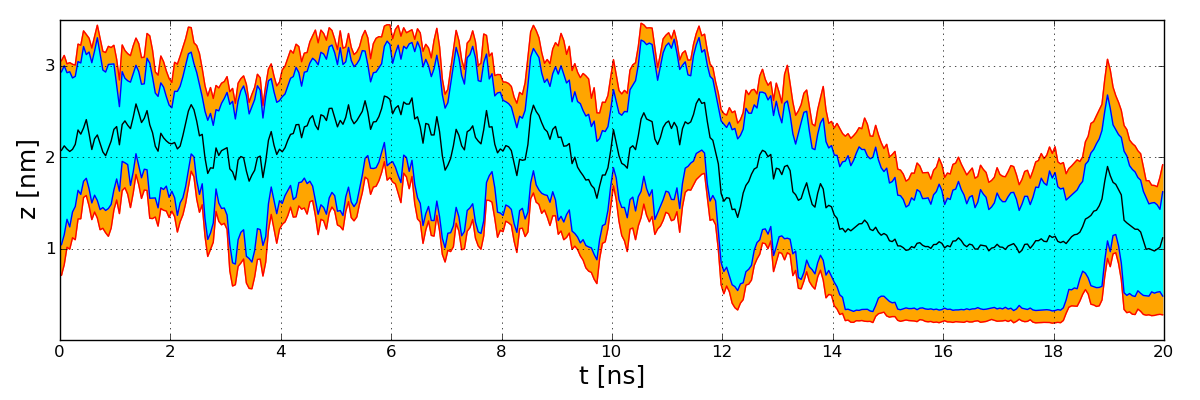}
\caption{An example of a free dynamic of 1L2Y in water above the surface of ZnS.
         Amino acids of the 1L2Y are capable of creating merely noncovalent bonds with the surface.
           The starting point corresponds to the CM placed at 2 nm above the surface.
         The lines are as in Figure \ref{cys_dyn}.}
\label{dis_1l2y}
\end{center}
\end{figure}

\begin{figure}[ht]
\begin{center}
\includegraphics[scale=0.5]{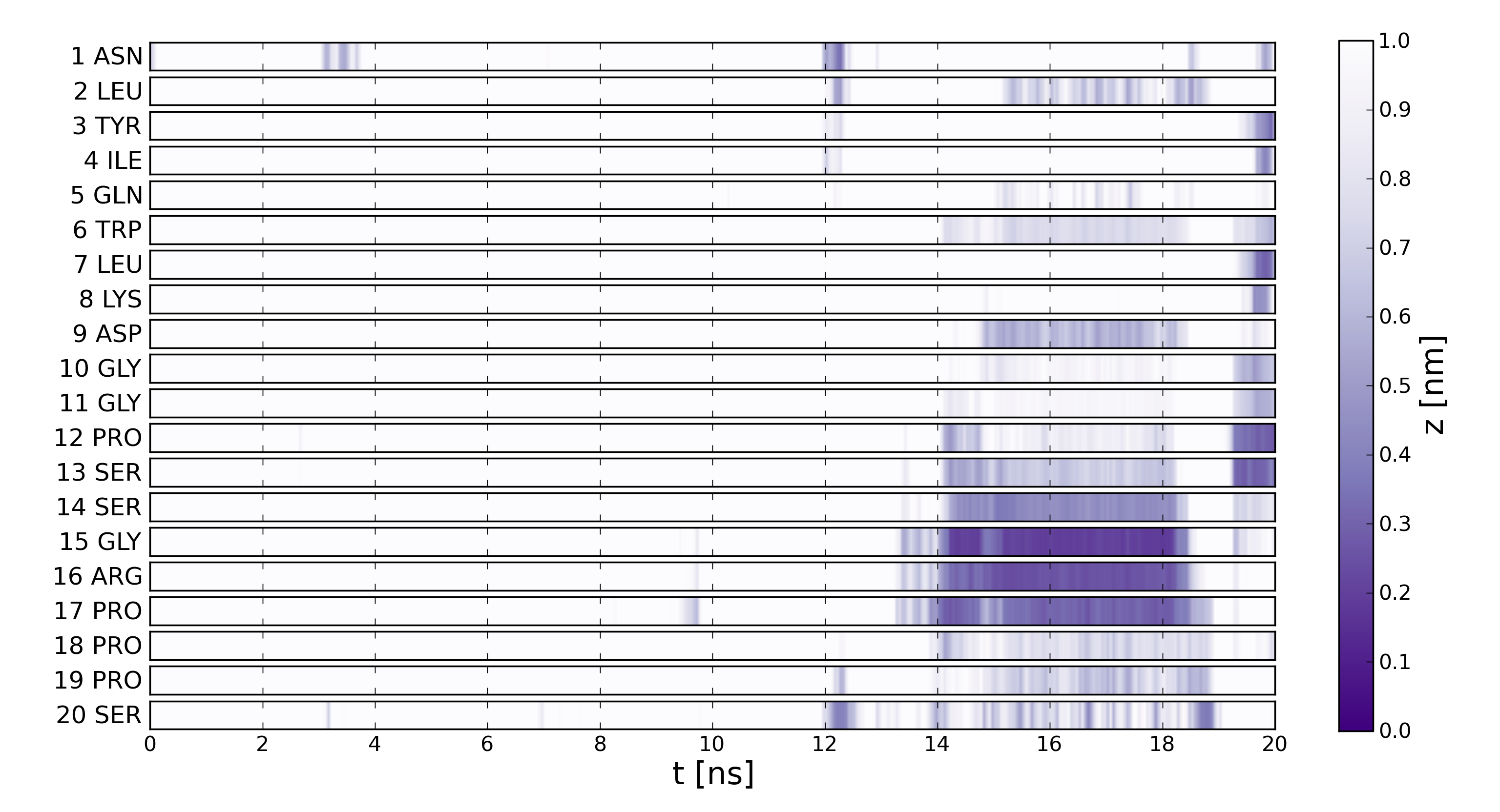}
\caption{The vertical positions of the lowest atoms of all amino acids in the 1L2Y 
         in the example of free dynamics shown on Fig. \ref{dis_1l2y}.}
\label{ads_1l2y}
\end{center}
\end{figure}

\begin{figure}[ht]
\begin{center}
\includegraphics[scale=0.5]{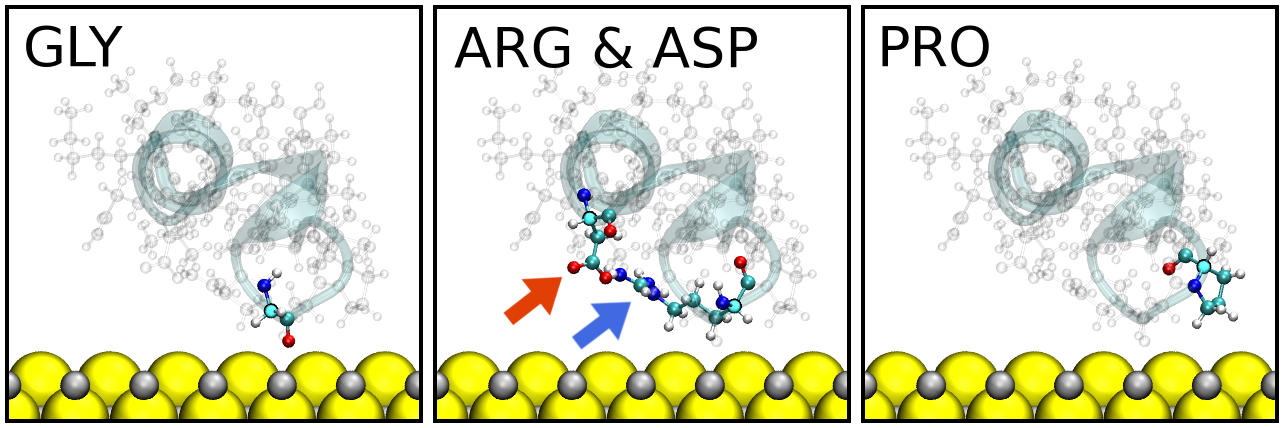}
  \caption{Snapshot of 1L2Y adsorbed at the surface of ZnS
           corresponding to the time of 16 ns in Fig. \ref{dis_1l2y} and \ref{ads_1l2y}.
           This instant is in the middle of the adsorption event.
            The snapshots focus on
           three  AAs that interact with the solid. In panel b), there is also
           ASP that interacts with ARG. 
           The red arrow points at the negatively charged carboxylate group of ASP
           and the blue one at the positively charged guanidinium group of ARG.
           The remaining AAs and the protain backbone
           are shown in the transparent representation.
           The molecules of water are omitted.}
  \label{con_aas_1l2y}
\end{center}
\end{figure}

In view of the generally weak values of $\epsilon$ shown in Table \ref{aas_val},
it is relevant to ask whether ZnS may lead to adsorption of a protein.
We first consider tryptophane cage (the structure code is 1L2Y) without any CYS.
This is a small protein as it comprises merely 20 AAs.
Our unrestrained simulation of 1L2Y in water solution 
shows that adsorption does take place but it is intermittent 
and the events of adsorption are brief.
One example of an adsorption event is shown in Fig. \ref{dis_1l2y}. 
Adsorption takes place between 14 and 18 ns and is driven
mainly by three  AAs:  GLY, ARG and PRO (Fig. \ref{ads_1l2y}). 
None of them has been classified by us as being attracted to ZnS: 
ARG is repeled and GLY and PRO are neutral.
However, the structure of the adsorbed protein suggests 
that the main driving force of adsorption comes from GLY, 
since the oxygen of its peptide bond substitutes one of water molecules 
in the direct interaction with Zn on the ZnS surface (Fig. \ref{con_aas_1l2y}). 
Nevertheless, also the hydrogen atoms of PRO cyclic stucture 
and of the ARG aliphatic chain 
interact with the sulfur atoms of the surface 
strengthening adsorption. 

This observation suggests that interactions between AAs within
a protein may affect conformations of the side groups and
generate higher affinity to the surface compared to single AAs. 
In the example of  Figs. \ref{dis_1l2y} and \ref{ads_1l2y}, the
positively charged guanidinium group of ARG  gets burried within
the surface of protein due to interactions with the negatively
charged ASP placed further away from the surface (Fig. \ref{con_aas_1l2y}). 
Thus, ARG hides the group that would otherwise cause repulsion, 
and simultaneously exposes the aliphatic chain of high affinity. 
This means that the binding energy for single AAs should be treated
rather as a suggestion  about the orders of magnitudes
-- when designing proteins --
than an exact guideline, 
especially when its value is low. 
Further analysis of the adsorbed protein shows 
that the interaction with the surface does not affect 
its structure and dynamics significantly. 

\begin{figure}[ht]
\begin{center}
\includegraphics[scale=0.5]{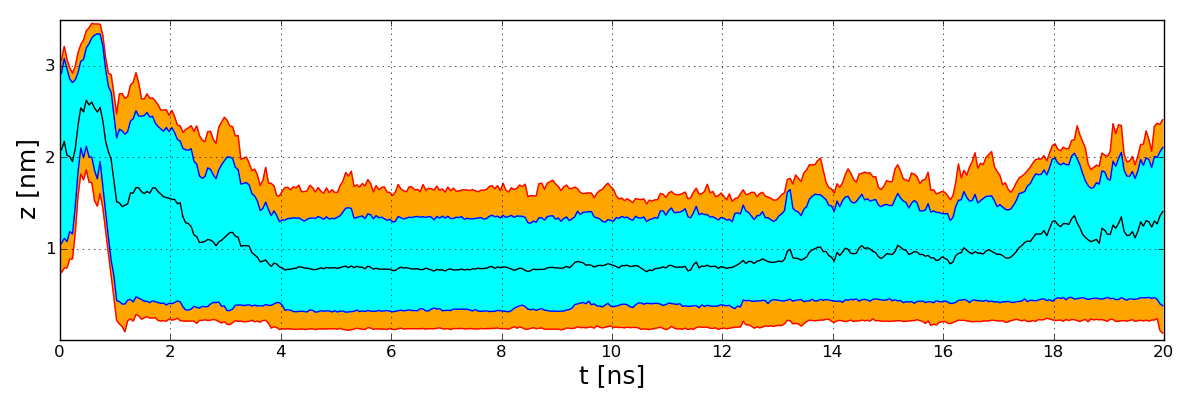}
\caption{An example of a free dynamic of m1L2Y in water above the surface of ZnS. 
         Cysteine attached to C-terminus is of the mCYS kind - it can form the disulfide bond with the surface.
         The lines are as in Fig. \ref{cys_dyn}.}
\label{dis_1l2yc}
\end{center}
\end{figure}

\begin{figure}[ht]
\begin{center}
\includegraphics[scale=0.5]{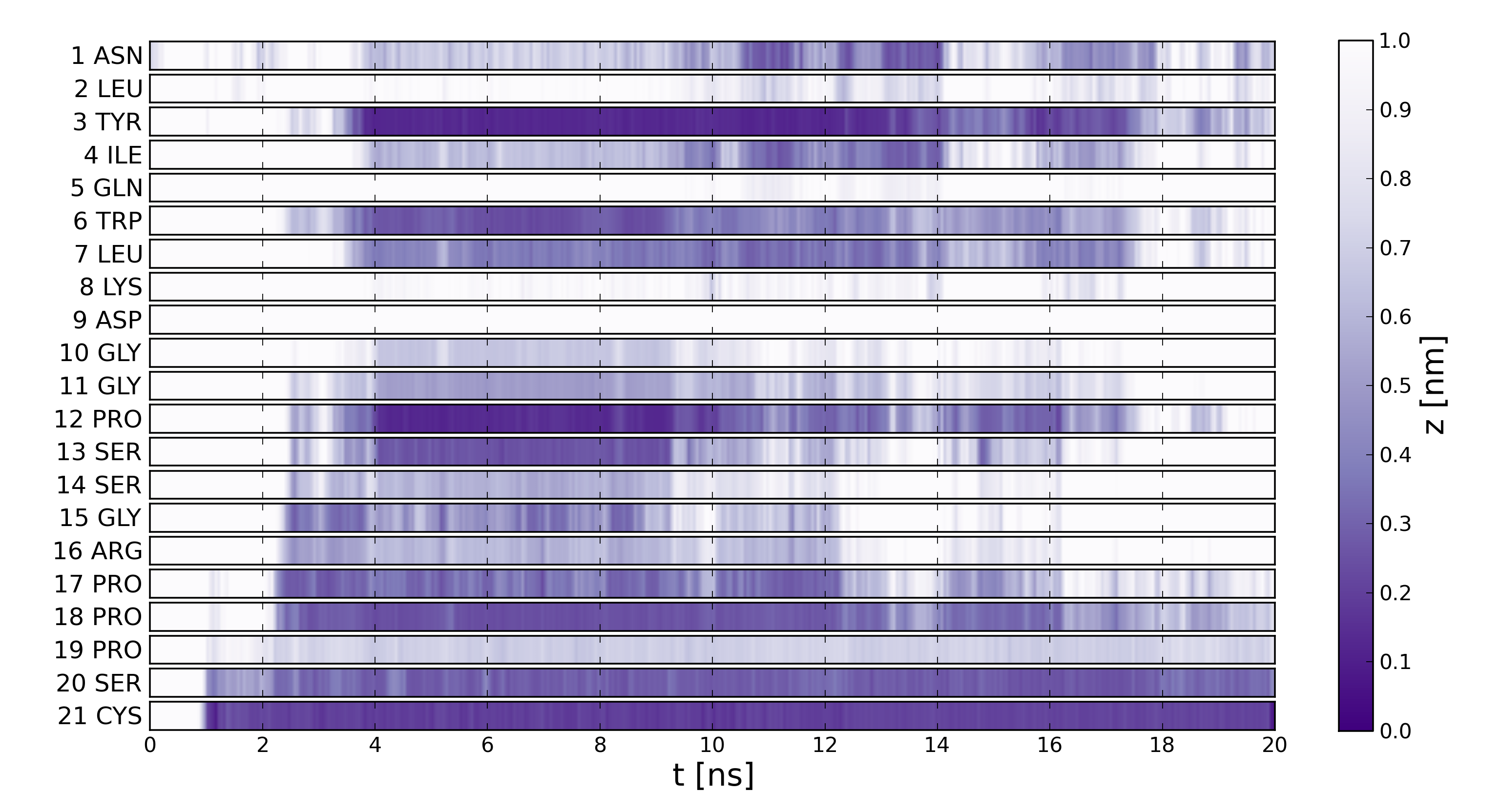}
\caption{The vertical positions of the lowest atoms of all amino acids in 1mL2Y 
         during the example of evolution shown on Fig. \ref{dis_1l2yc}.}
\label{ads_1l2yc}
\end{center}
\end{figure}

\begin{figure}[ht]
\begin{center}
\includegraphics[scale=0.4]{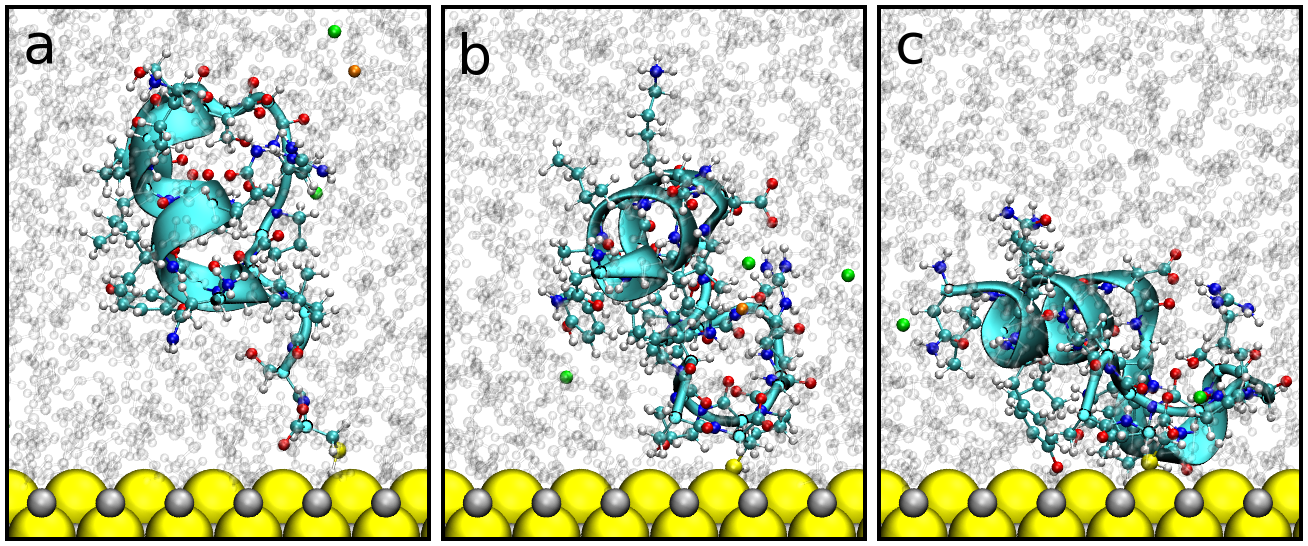}
\end{center}
  \caption{m1L2Y during adsorption to the surface of ZnS in water solution.
           The snapshots correspond to time of 1 ns (a), 3 ns (b) and 5 ns (c) 
          in runs illustrated in Figs. \ref{dis_1l2yc} and \ref{ads_1l2yc}.
           The isolated spheres show some of the ions: Cl$^{-}$ in green and Na$^{+}$ in orange.}
  \label{ads_1l2yc_pics}
\end{figure}

\begin{figure}[ht]
\begin{center}
\includegraphics[scale=0.45]{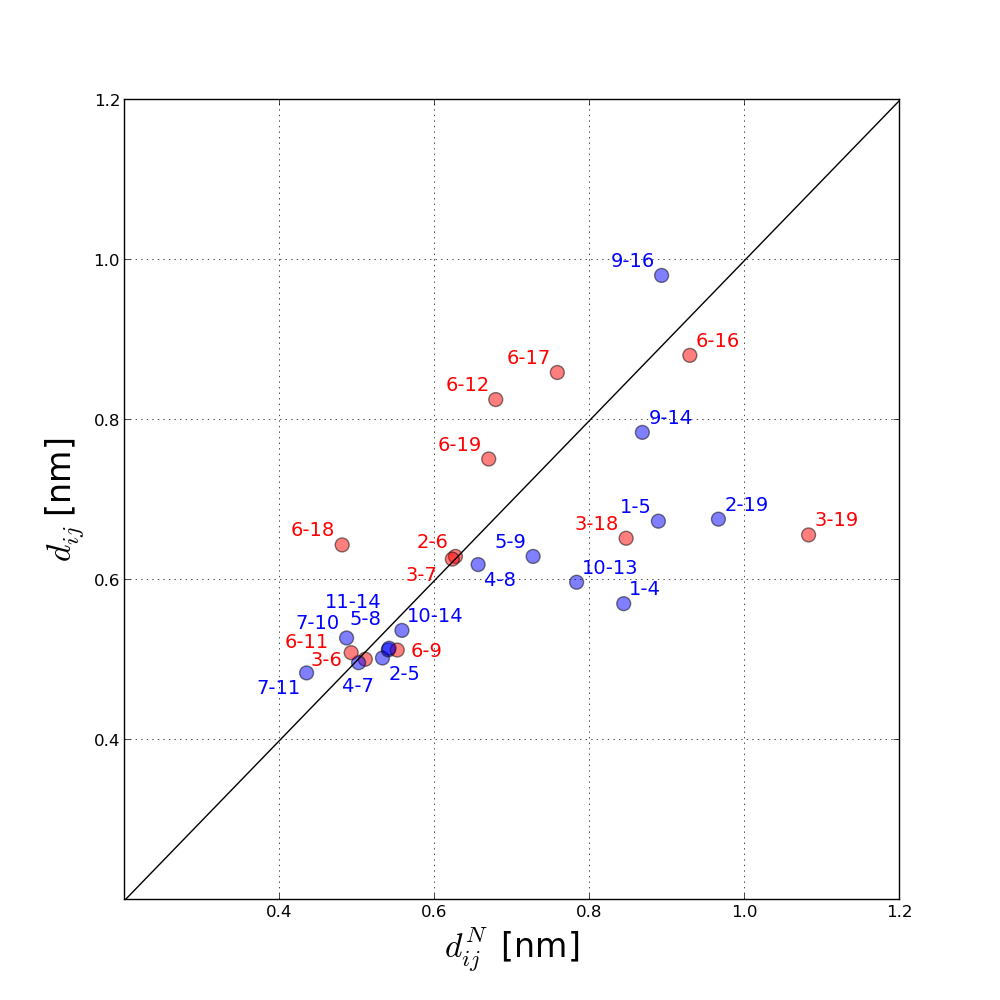}
\caption{Dynamics of m1L2Y during a binding event to the ZnS surface. 
         The event engages five AAs: at sites 3, 6, 12, 18 and 20. 
         The figure displays average distances in the native contacts in the bound state 
         against the same distances in the bulk water.
         The red color is used if at least one of the AA 
        in the contact is involved in the binding.
         The averages are over the time interval between 4 and 9 ns of the 
dynamic shown in Fig. \ref{dis_1l2yc} and \ref{ads_1l2yc}.}
\label{proncd}
\end{center}
\end{figure}

In order to investigate the role of mCYS in a protein adsorption 
we modify 1L2Y by attaching cysteine to its C-terminus. 
This modified 1L2Y is denoted here as m1L2Y. We have generated eight 
trajectories of m1L2Y evolving near the surface of ZnS in water solution. 
We observe occurrence of adsorption for about 54\% of 
entire time (an example is shown in Fig. \ref{dis_1l2yc})
and not merely 13\% obtained for 1L2Y.
A big portion of it (44 \%) is due to the interactions with mCYS.
Once m1L2Y makes a contact with the surface 
it generally stays adsorbed for the rest of the simulation.
It detaches only occasionally.
Furthermore, our analysis of the covalent bond dynamics 
reveals that its breakage is about 10 times rarer 
when the mCYS is a part of the protein rather than in isolation. 
This means that the presence of mCYS not only extends the adsorption 
time of the protein 
but it also extends the duration of the bond with mCYS. 
The latter can be a result of both  the additional interactions with
other AAs with the surface (see Fig. \ref{ads_1l2yc} and \ref{ads_1l2yc_pics})
and the stabilization of covalent bond by the protein.
The enhanced stabilization comes from the larger moment of
inertia of the protein and thus smaller influence of of water.
Adsorption of m1L2Y to the surface of ZnS does not lead any significant 
deformation of the protein (see Fig. \ref{ads_1l2yc_pics}). This is 
similar to the case of the ZnO surface \cite{nawrocki_2013} 
but opposite to the case of the polar mica \cite{starzyk_2011,starzyk_2013}. 
The adsorption changes average distances in the native contacts
by no more than 0.4 nm.
This feature is illustrated in Fig. \ref{proncd} where average
contact distances are plotted against the native distances in the
contacts -- the outlying contact 3-19 is close to the surface.
The native contacts are defined through the overlap of enlarged
van der Waals spheres associated with the heavy atoms, as used
in the dynamics of the Go-like models \cite{tsai_1999,sikora_2009}.

An alternative way to characterize the propensity for adsorption
of a protein is through the PMF for the protein.
We determine it by using the umbrella sampling, $i.e.$ through 
the same procedure that we have applied for the capped AAs. 
We make independent runs that start by pulling 1L2Y 
towards the surface from  various initial orientations.
We note
that the orientations can switch freely because the
the protein is constrained  only at its center of mass. 
Thus, the simulations lasting for a sufficiently long time 
should yield exactly the same PMF. However, achieving such a time
scale and generating appropriate statistics is difficult even
for the protein of merely 20 AAs. 
Here, we focus instead on a certain cutoff time range in which
the the results are biased by the nature of the initial orientation. 

\begin{figure}[ht]
\begin{center}
\includegraphics[scale=0.5]{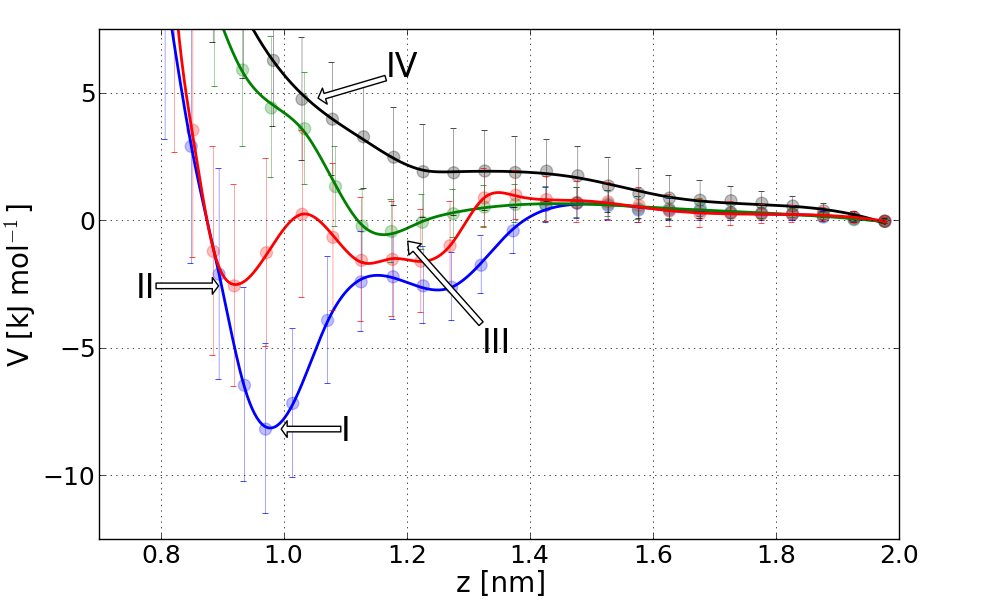}
\caption{Potential of the mean force for protein 1L2Y in water
         as a function of the distance $z$ above a surface of ZnS. 
         Each of the four plots relates to independent procedure of pulling and 
         umbrella sampling dynamic
         that start from different initial orientations of the protein.}
\label{pmf_1l2y}
\end{center}
\end{figure}

Fig. \ref{pmf_1l2y}  shows four resulting plots of $V(z)$.
The PMF denoted as I has a minimum at 8.1 kJ/mol.
This energy is almost twice the $\epsilon$ corresponding
to the strongest binding single AA (THR -- 4.28 kJ/mol)
if one excludes CYS. This suggests that various AAs combine
to build strength of binding.
On the other hand, the PMF denoted as IV corresponds to 
repulsion at all distances.
The remaining PMFs, denoted as II and III, correspond to
weak and very weak binding respectively.

\begin{figure}[ht]
\begin{center}
\includegraphics[scale=0.3]{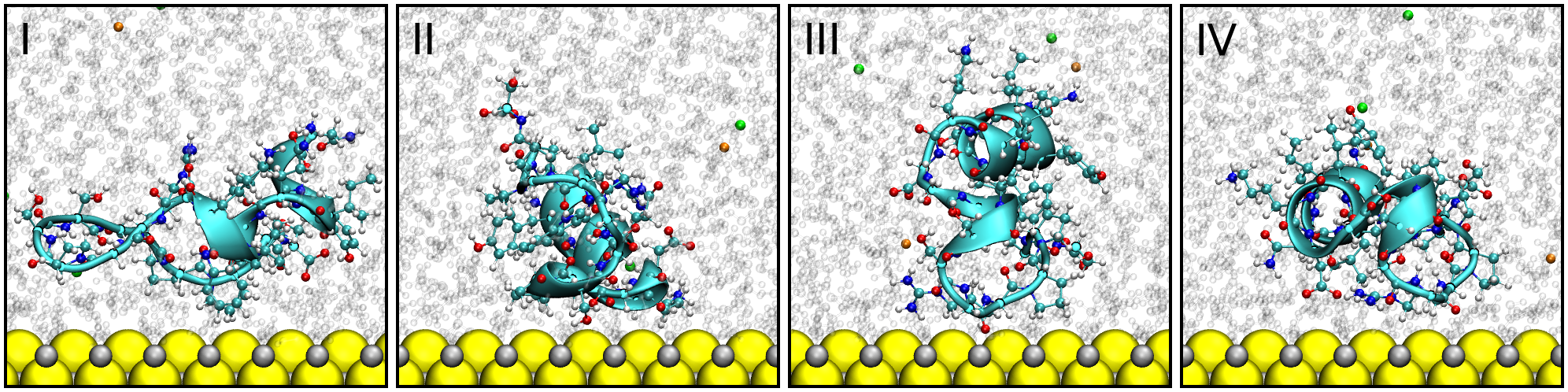}
\end{center}
  \caption{The optimal conformations of 1L2Y protein at the ZnS (110) surface
           corresponding to the individual PMF curves of Fig. \ref{pmf_1l2y} 
           as labeled by the Roman numerals.}
  \label{pro_umb}
\end{figure}

\begin{figure}[ht]
\begin{center}
\includegraphics[scale=0.5]{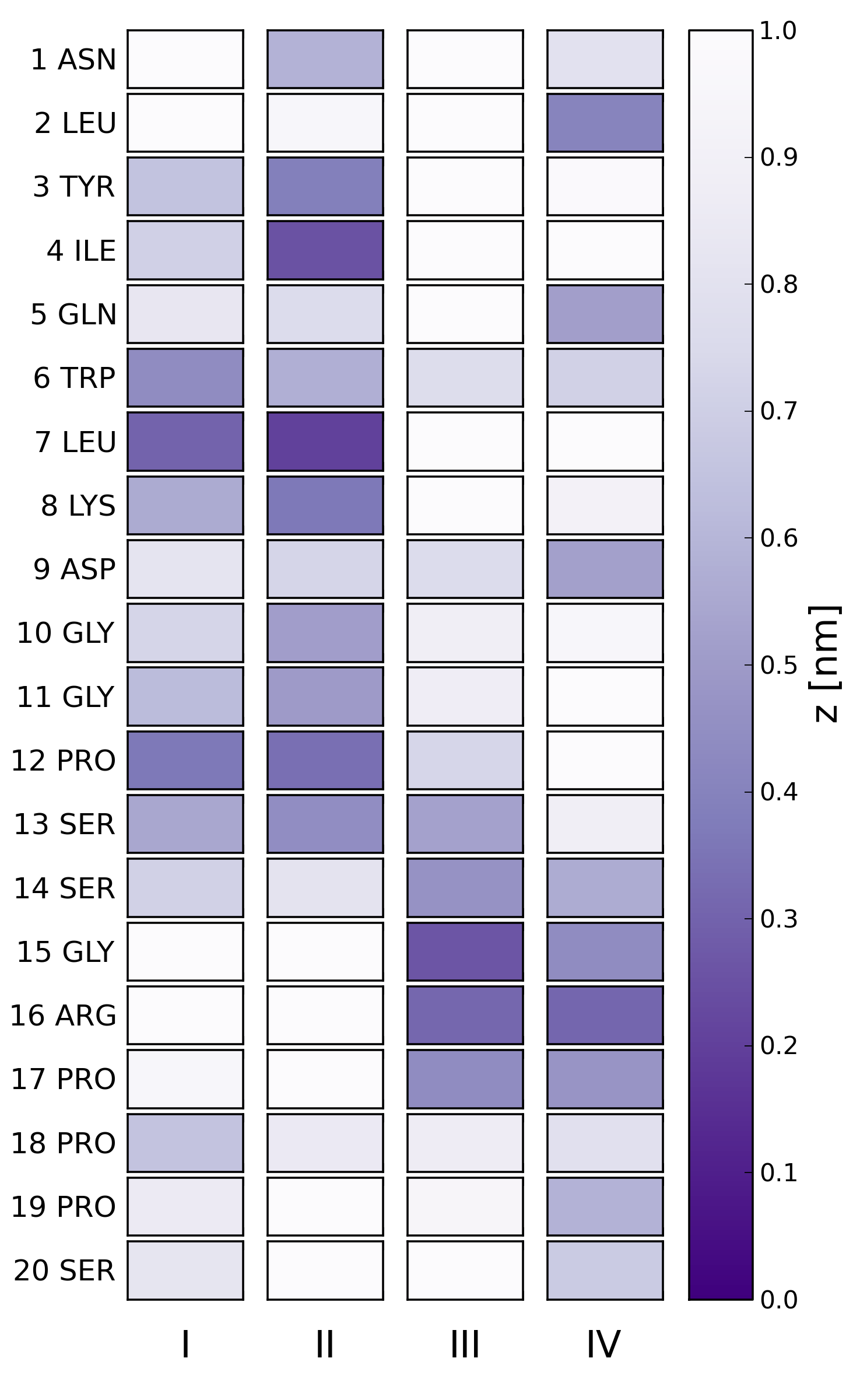}
\caption{The average vertical positions of the lowest atoms belonging to the individual AAs in 1L2Y; 
         the optimal points of the PMF denoted in Fig. \ref{pmf_1l2y} by the corresponding Roman numerals.} 
\label{ads_1l2y_pmf}
\end{center}
\end{figure}

Fig. \ref{pro_umb} shows examples of 1L2Y conformations 
near the minimal points in the $V(z)$ of Fig. \ref{pmf_1l2y}
(or at 1 nm for PMF IV). They are all quite distinct.
In order to determine which AAs interact with the surface in each of the cases 
we calculate the average vertical positions of the lowest atoms of all AAs 
in the adsorbed protein. The results for the four
potentials are displayed in Fig. \ref{ads_1l2y_pmf}. 
For I and II, the relevant AAs that are coupled to the solid are
around LEU7 and PRO12. The latter has been already recognized as
important in the free-evolution runs (Fig.  \ref{con_aas_1l2y}). 
The main difference between I and II is that II also involves binding
of ILE4 -- and yet the resulting $V(z)$ is observed to be higher.
In case III, binding moves up the sequence to the vicinity of GLY15.
There is no effective binding in case IV but the AA which is closest
to the solid is ARG16.

\section{Summary}

The main result of our studies is that the energies of binding
of AAs to ZnS (110) surface are similar in magnitude to those to the
four faces of ZnO. However the aspects of specificity are different.
For instance, THR which binds to ZnS the strongest does not bind to
three faces of ZnO and binds only very weakly to the fourth one
(to (000$\bar{1}$)-Zn). This is because the crystal structures
are distinct and the water density profiles are quite different.
The binding strength is comparable to that provided by hydrogen
bonds in proteins. The exception is cysteine for which formation
of the disulfide bond with an atom of S in the solid increases
the binding strength by an order of magnitude. The covalent binding
also increases the duration of the adsorption events in the model.
This is in a qualitative agreement with experimental studies of peptides 
that bind to the ZnS particles.  For instance,
Pinaud $et$ $al.$ \cite{pinaud_2004} have shown that the replacement 
CYS by ALA results in formation of particles that are
insoluble in water because the coverage by the peptides
is reduced significantly.
Gondikas et $al.$ \cite{gondikas_2010} have observed
that SER does not adsorb in appreciable amounts to the particle surfaces 
but CYS does. Thus SER, unlike CYS, does not appear to alter effective
surface properties of ZnS.
A further exploration of the protein-solid interactions appears to be 
important for making progress in bio-nanotechnology.

{\bf Acknowledgments}
Discussions with D. Elbaum, J. Grzyb and B. R\'o\.zycki are warmly appreciated. 
This work has been supported by 
the Polish National Science Centre Grants No. 2011/01/B/ST3/02190 (MC and GN) 
and N N202 130039 (GN) 
as well as by the European Union within European Regional Development Fund, 
through Innovative Economy grant (POIG.01.01.02-00-008/08). 
The local computer resources were financed by the European Regional Development Fund 
under the Operational Programme Innovative Economy NanoFun POIG.02.02.00-00-025/09. 
We appreciate help of A. Koli{\'n}ski and A. Liwo in providing additional 
computer resources such as at the Academic Computer Center in Gda\'nsk.

\clearpage

\begin{table}[ht]
\begin{tabular}{c|c|c}
& $\epsilon$ & $\sigma$ \\
& [kJ/mol] & [nm] \\
\hline
ASP &    $-$  &  $-$  \\
GLU &    $-$  &  $-$  \\
CYS &    1.77  &  0.43  \\
\textcolor{gray}{mCYS} & \textcolor{gray}{98.35} & \textcolor{gray}{0.40} \\
ASN &    0.79  &  0.50  \\
PHE &    1.52  &  0.51  \\
THR &    4.28  &  0.44  \\
TYR &    3.06  &  0.53  \\
GLN &    0.58  &  0.50  \\
SER &    1.95  &  0.43  \\
MET &    $-$  &  $-$  \\
TRP &    0.55  &  0.62  \\
VAL &    2.11  &  0.45  \\
LEU &    1.45  &  0.49  \\
ILE &    1.77  &  0.50  \\
GLY &    0.32  &  0.45  \\
ALA &    0.28  &  0.47  \\
PRO &    0.41  &  0.56  \\
\textcolor{gray}{HIE} & \textcolor{gray}{1.19} & \textcolor{gray}{0.55} \\
\textcolor{gray}{HID} & \textcolor{gray}{3.06} & \textcolor{gray}{0.45} \\
\textcolor{gray}{HIP} & \textcolor{gray}{0.38} & \textcolor{gray}{0.80} \\
LYS &    $-$  &  $-$  \\
ARG &    $-$  &  $-$  \\
\hline
average and dispersion without mCYS & \textbf{1.49 $\pm$ 1.07} & \textbf{0.50 $\pm$ 0.06} \\
average and dispersion with mCYS & \textbf{7.55 $\pm$ 23.47} & \textbf{0.49 $\pm$ 0.06} \\
\end{tabular}
\caption{Values of the binding energy, $\epsilon$,  and of the bond length, 
$\sigma$, between the center of mass of an AA and the ZnS surface  as determined
through the umbrella sampling method.
The symbol -- signifies a non-binding situation. 
Non-binding situations do not contribute to the average and dispersion of 
$\epsilon$ and $\sigma$ that are listed at the bottom. 
Results corresponding to the various forms of histidine are first averaged to form one entry.}
\label{aas_val}
\end{table}

\clearpage

\clearpage

\clearpage

\end{document}